\documentclass[cits]{PoS}

\title{Characterization of a Spherical Proportional Counter in argon-based mixtures}

\ShortTitle{The Spherical Proportional Counter}

\author{\speaker{F.J.~Iguaz}, A.~Rodr\'iguez, J.F.~Castel and I.G.~Irastorza\\
        Laboratorio de F\'isica Nuclear y Astropart\'iculas, Universidad de Zaragoza, Spain.\\
        E-mail: \email{iguaz@cern.ch}}

\abstract{The Spherical Proportional Counter is a novel type of radiation detector, with a low energy threshold
(typically below 100 eV) and good energy resolution. This detector is being developed by the network NEWS,
which includes several applications. We can name between many others Dark Matter searches,
low level radon and neutron counting or low energy neutrino detection from supernovas or
nuclear reactors via neutrino-nucleus elastic scattering.
In this context, this works will present the characterization of a spherical detector of 1 meter diameter
using two argon-based mixtures (with methane and isobutane) and for gas pressures between 50 and 1250 mbar.
In each case, the energy resolution shows its best value in a wide range of gains,
limited by the ballistic effect at low gains and by ion-backflow at high gains.
Moreover, the best energy resolution shows a degradation with pressure.
These effects will be discussed in terms of gas avalanche properties.
Finally, the effect of an electrical field corrector in the homogenity of the gain
and the energy threshold measured in our setup will be also discussed.}

\FullConference{Technology and Instrumentation in Particle Physics 2014\\
		 2-6 June, 2014\\
		 Amsterdam, the Netherlands}

\begin{document}
\section{Introduction}
An active field in instrumentation for Particle Physics is the development of massive detectors with low background
levels and low energy threshold. The range of applications is diverse: the search of dark matter in the form of
Weakly Interacting Massive Particles (WIMPs) \cite{Baudis:2012lb} or axions \cite{Wantz:2010},
the detection of low-energy neutrinos, radon and neutron counting.
In this context, the Spherical Proportional Counter (SPC) was proposed in \cite{Giomataris:2008ig}
as a candidate fulfilling all the requirements: a single readout channel reading a big gas volume (and mass);
a potentially low intrinsic background level with a good selection of radiopure materials;
and a high signal-to-noise ratio due to its low capacitance, which is proportional to the radius
of the central electrode and does not depend on the vessel size.
Indeed, previous works have shown that an energy threshold as low as 100 eV \cite{Bougamont:2012eb, Dastgheibi:2014ad}
is feasible, keeping a good energy resolution (10\% FWHM at 22.1 keV).

\begin{figure}[htb!]
\centering
\includegraphics[width=75mm]{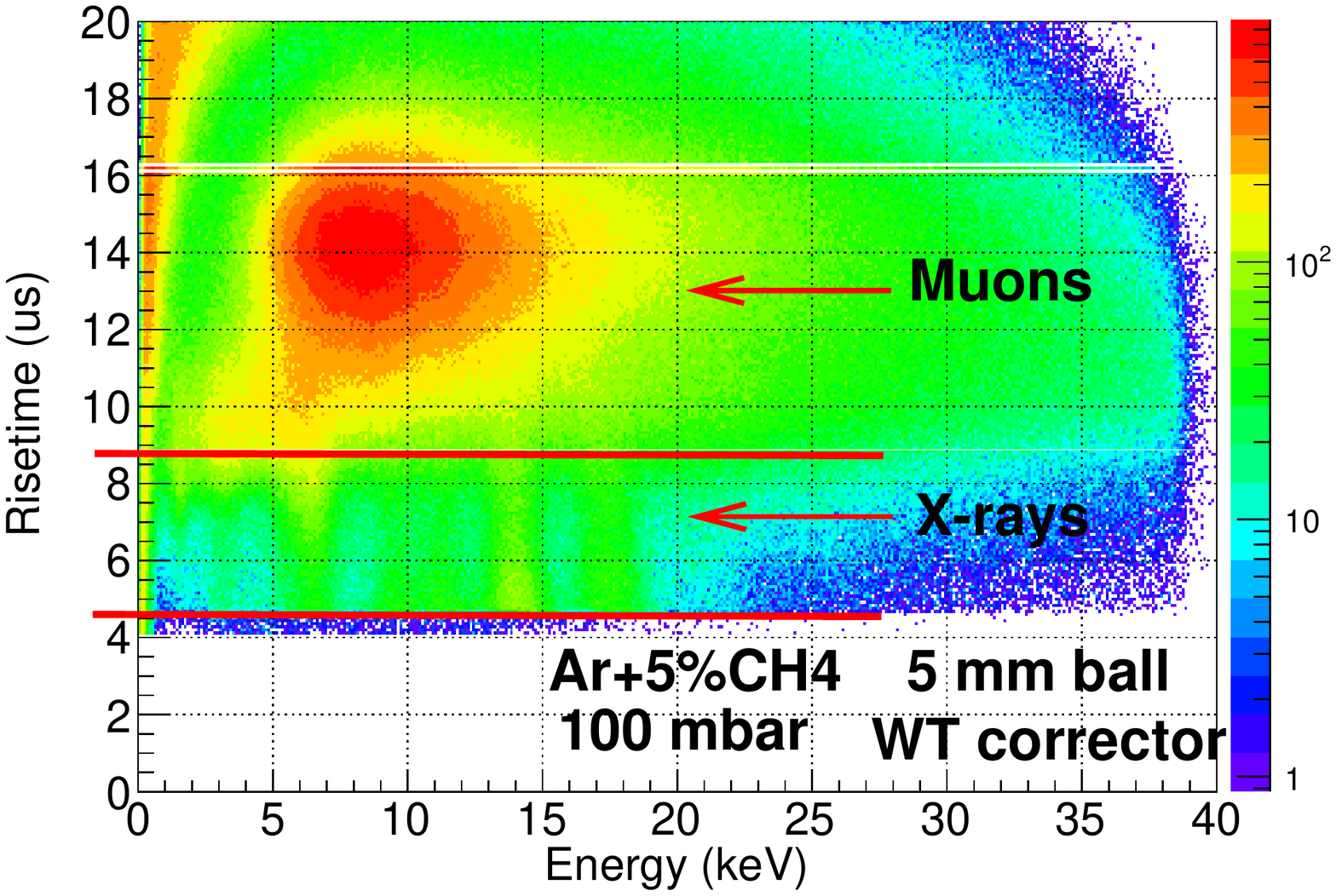}
\includegraphics[width=75mm]{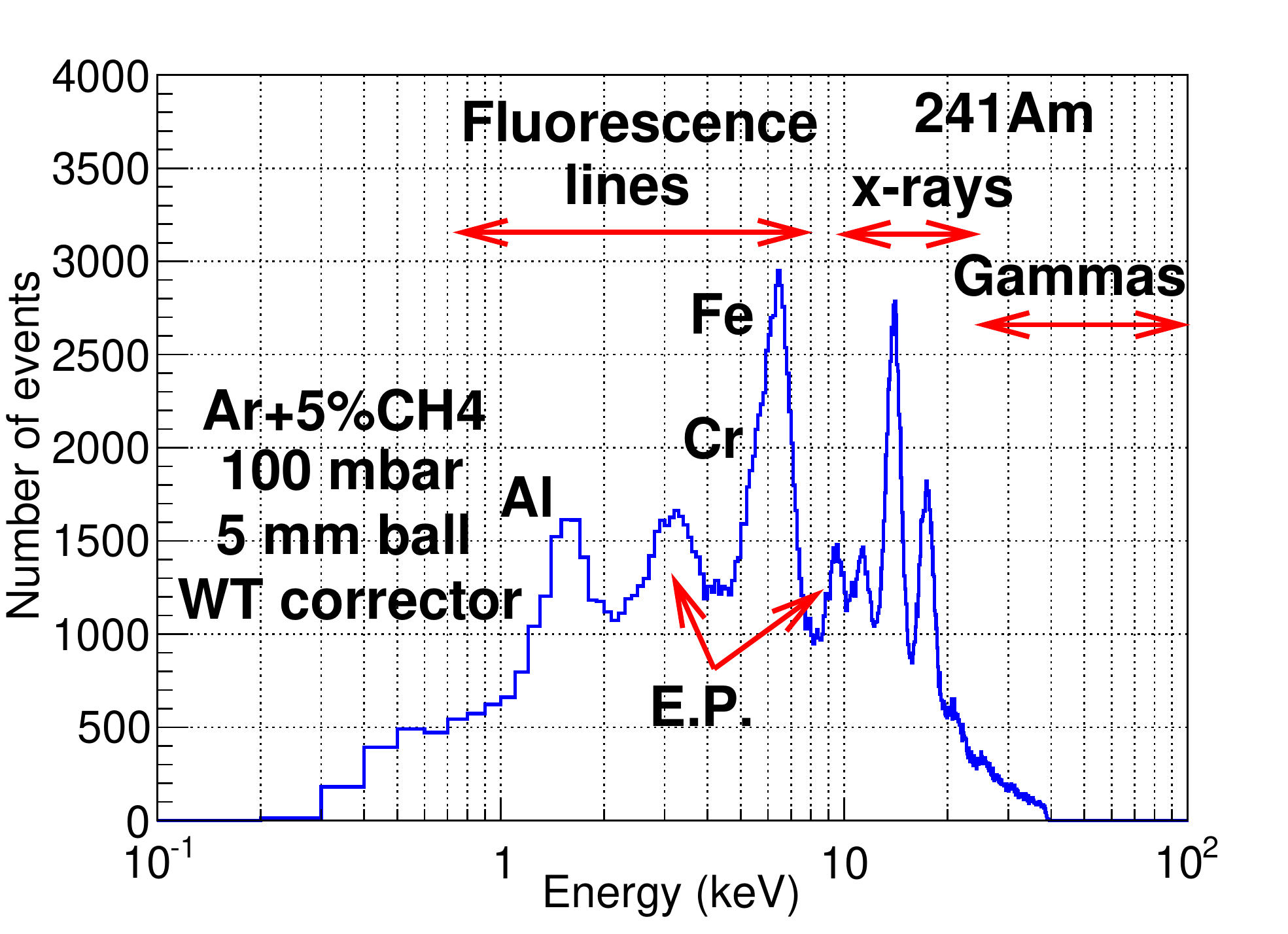}
\caption{Left: Dependence of the risetime with the pulse amplitude in Ar+5\% methane at 100 mbar,
obtained irradiating our SPC with a $^{241}$Am source whose 5.5 MeV alphas have been blocked by an aluminum foil.
Muons show an event distribution centered at 8 keV and 14 $\mu$s. Some muons events
(see band below 3 keV at long risetime values) are not well analized as their risetime is longer than the analysis range.
The x-rays from the source (at 11.9, 13.9, 17.7 and 20.8 keV) create different bands in risetime
while the fluorescence lines emitted from the vessel (chromium at 5.5 keV and iron at 6.4 keV)
and the aluminum foil (at 1.5 keV) form three spots at long risetime.
The red lines delimit the fiducial volume for point-like events.
Right: Energy spectrum after having selected the events in the fiducial volume. The Argon escape peaks (E.P.) of
both fluorescence lines and $^{241}$Am x-rays lines are also present in the spectrum.}
\label{fig:RTDisc}
\end{figure}

\medskip
Even if there is only one channel, the pulse risetime can be used to discriminate complicate topologies like muons
from point-like events like x-rays; or to set a fiducial volume, as the risetime depends on the radial position
where the energy was deposited by gas diffusion. As an example of this feature, we present in figure \ref{fig:RTDisc}
(left) the dependence of the risetime with the pulse amplitude for our SPC irradiated by a $^{241}$Am source
covered by an aluminum foil. As the mean free path of photons is longer than the sphere's radius
at low pressures, an x-ray may deposit its energy at any distance. As a result, it will create
a distribution in risetime for a fixed amplitude in the 2D plot. This fact happens for the x-rays
of the source (at 11.9, 13.9, 17.7 and 20.8 keV). In contrast, the fluorescence lines emitted
from the stainless steel vessel (chromium at 5.5 keV and iron at 6.4 keV) and the foil (aluminum at 1.5 keV)
form three spots at long risetimes. If we select the events whose risetime ranges
in 4.5-8.7 $\mu$s (red lines), we obtain the energy spectrum of figure \ref{fig:RTDisc} (right).

\medskip
The advantages of SPCs have motivated several feasibility studies \cite{Dastgheibi:2014ad,
Gerbier:2014gg, Vergados:2009jdv, Galan:2013jg, Giomataris:2010ig} and the groups interested in this technology have created
the network NEWS (New Experiments With Spheres), which is now formed by institutes of France, Greece, China and Spain.
In this context, the scalability of high gains and good energy resolutions to higher masses (pressures) is an open question,
which is a key-point to increase the sensitivity of the detector to any particular application.
This work tries to give a first answer, presenting the characterization of a SPC
in two argon-based mixtures (with methane and isobutane) and for gas pressures up to 1250 mbar.
The setup, the data-taking procedure and the analysis are described in detail in section 2.
We also present the two configurations of the cental rod (without and with a field corrector),
whose main results will be respectively detailed in sections 3 and 4.
Finally, we finish in section 5 with some conclusions and an outlook.

\section{Detector description}
\begin{figure}[htb!]
\centering
\includegraphics[width=120mm]{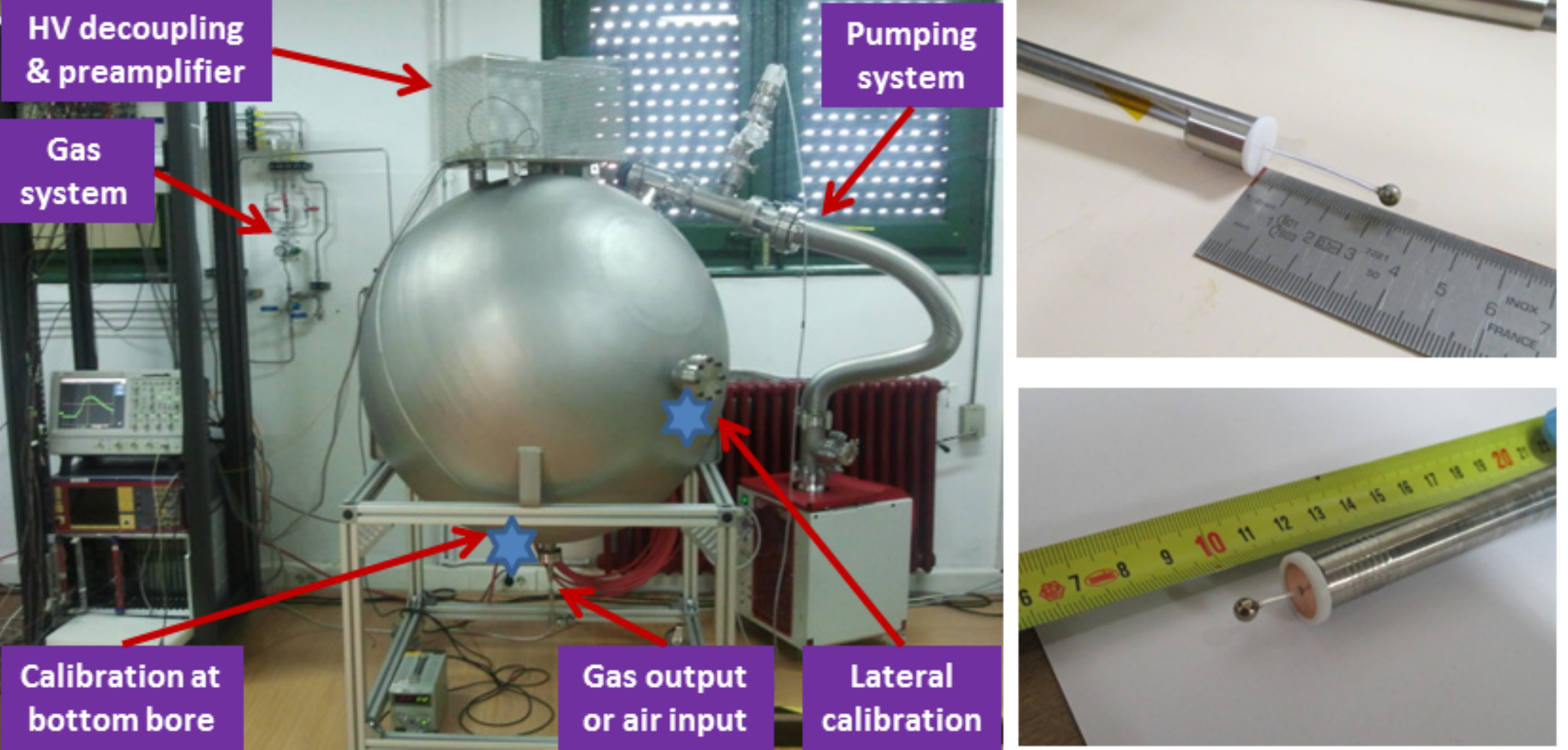}
\caption{Left: A general view of the spherical proportional counter characterized in this work
and the auxiliar facilities: the gas system, the pumping system and the in/output valves. The blue stars
indicate the points were the cadmium source was installed, as described in the text.
Right: the two configurations tested in this setup: a central ball without (top)
and with (bottom) an umbrella field corrector.}
\label{fig:Setup}
\end{figure}

The detector consists of a large spherical stainless steel vessel 1.0 meter in diameter
and a small metallic ball 5 mm in diameter kept at the center by a stainless steel rod.
The radiation from the calibration source or other sources (like muons and gammas) ionize the gas.
As the central ball is set to a positive voltage and the vessel to ground, electrons drift to the ball
where an intense electric field amplifies the charge. The charge movement induces a signal at the electrode,
which is extracted by a teflon-based high voltage feed-through from the vessel.
The signal is decoupled from the high voltage (powered by a CAEN N1471HA module) by a capacitance and
then fed into a CANBERRA preamplifier, which is acquired by a Tektronix DS5054B oscilloscope.
A view of the setup is shown in figure \ref{fig:Setup} (left), while on right caption,
we show the two configurations used in these tests: in the first one the ball is situated at 30 mm
of the rod, covered by a teflon cap; in the second one, the ball is 20 mm away from a copper plate, which
works as a field corrector (generally called ``umbrella'') and is isolated from the rod by a teflon cassette.

\medskip
The detector operates in seal mode, i.e., the vessel is first pumped out and then filled with an appropriate gas
at a pressure from few tens of mbar up to 2 bar. Two mixtures have been studied: Ar+5\% methane for pressures between
50 and 1250 mbar and Ar+2\% isobutane for pressures between 100 and 400 mbar.
No gas filter has been used in this work. The pumping lasted at least one night
to reduce the outgassing rate from inner vessel's components below $10^{-6}$ mbar l/sec.
This value allows the operation of the detector during a week without any visible degradation.

\medskip
The results presented here were obtained irradiating the SPC by a $^{109}$Cd source respectively
situated inside the vessel at the bottom part for the rod without the field corrector (figure \ref{fig:Setup}),
and at the lateral bore for the second configuration. In the first case, we observed
that the energy resolution was poor ($\approx$30\% FWHM at 22.1 keV) when calibrating from the lateral bore.
In contrast, values were good ($\approx$10\% FWHM) when the SPC was calibrated from the bottom part.
We concluded that the field was not homogeneous and that a field corrector was needed,
as suggested in \cite{Giomataris:2008ig}. However, as described in section 4, the field became homogeneous only when
the umbrella was set to negative values. This fact disagrees with the conclusions of the former publication. A simulation
of the electric field is foreseen to evaluate the effect of the field corrector in our setup.

\medskip
In a first step, signals acquired by the oscilloscope are smoothed. The first derivate is then calculated
to define a temporal range where the Pulse Shape Analysis (PSA) is applied. This range is useful to avoid errors
produced by pile-up events, unstable baselines or noise. Finally, the PSA calculates some pulse features like
the baseline, risetime, or amplitude. As shown in figure \ref{fig:RTDisc}, the risetime allows to discriminate
x-rays from muons setting an upper limit.

\begin{figure}[htb!]
\centering
\includegraphics[width=75mm]{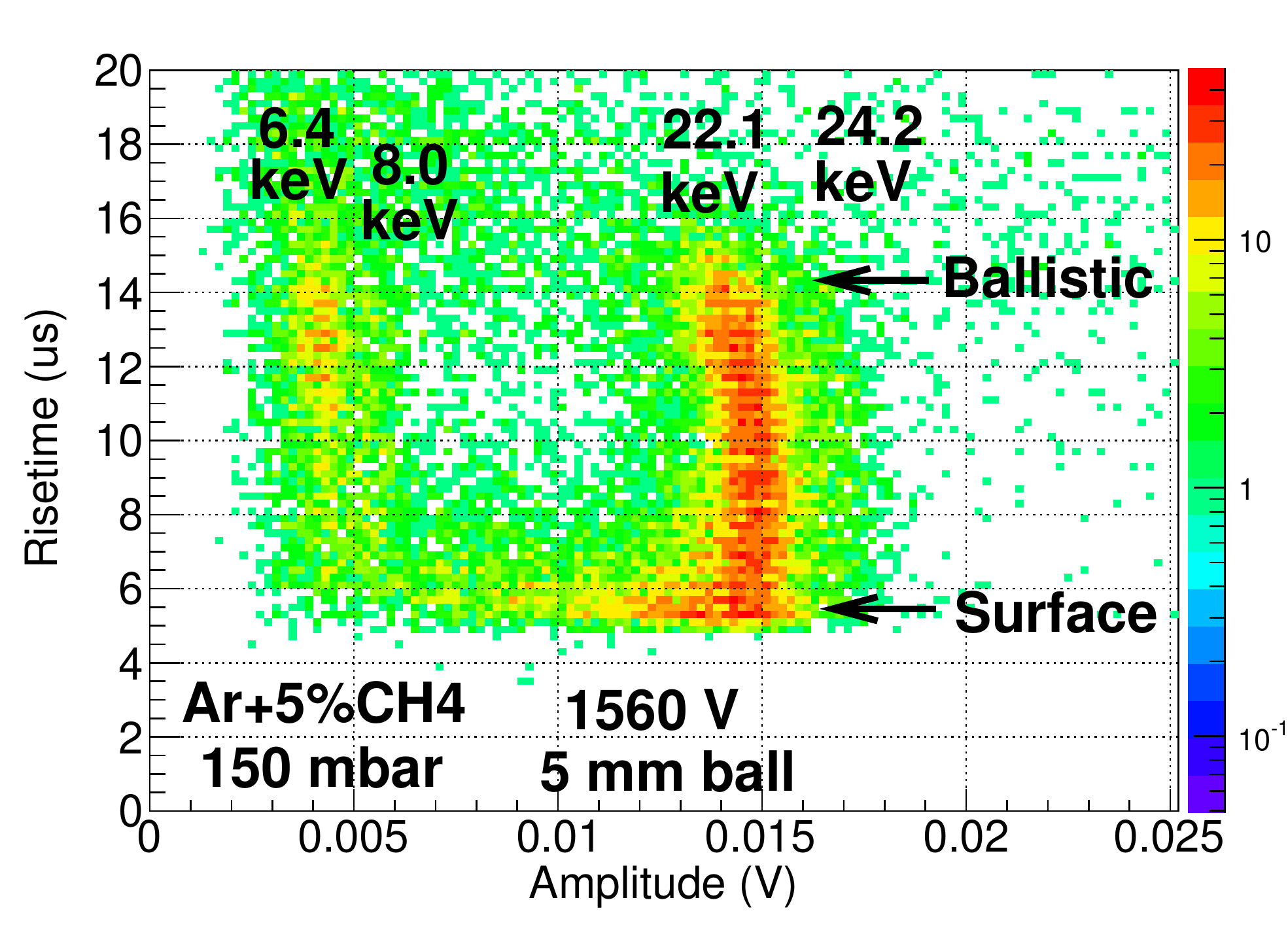}
\includegraphics[width=75mm]{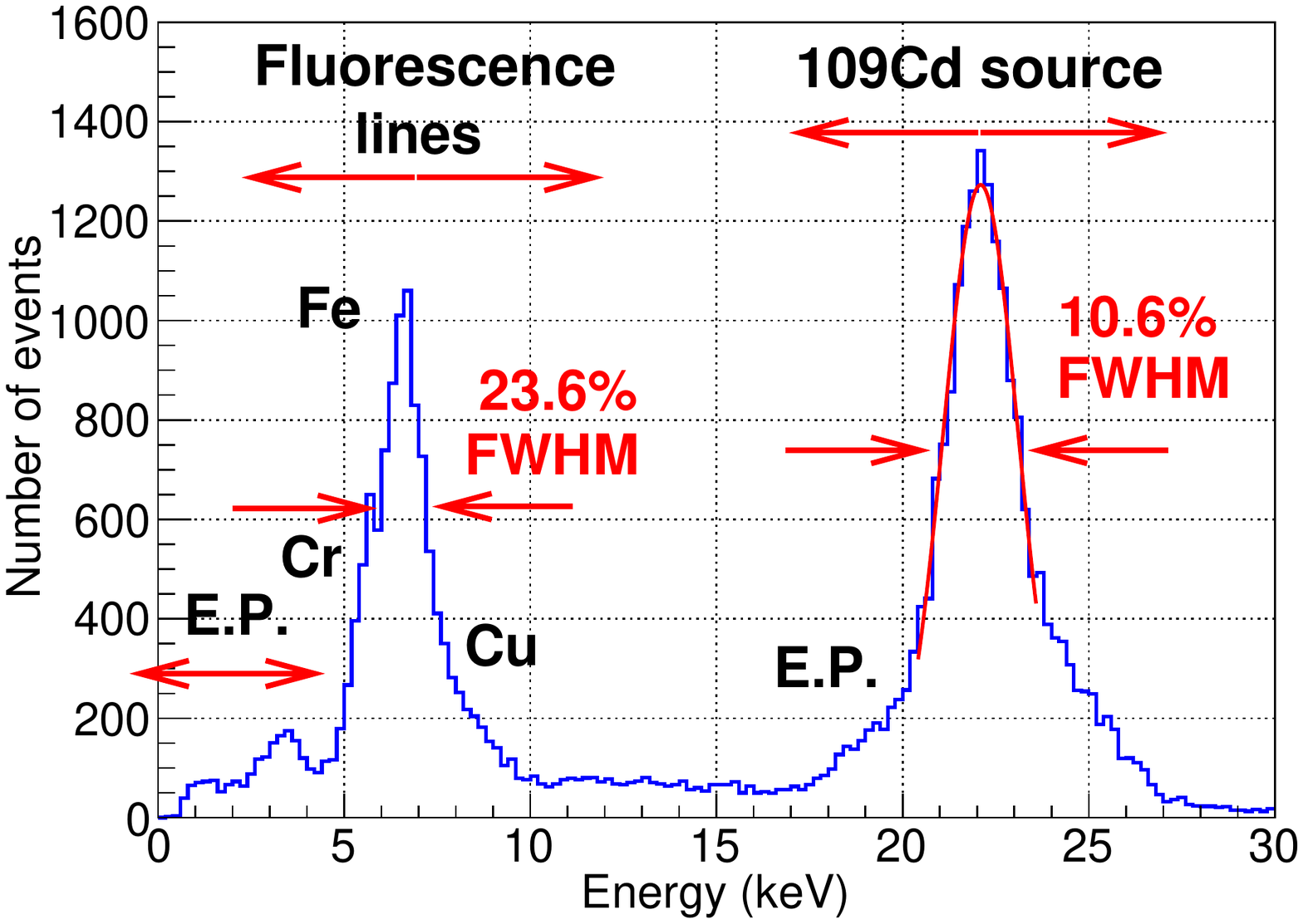}
\caption{Left: Dependence of the risetime with the pulse amplitude for a low voltage case in Ar+5\% methane at 150 mbar,
obtained irradiating the SPC with a $^{109}$Cd source.
Right: A typical energy spectrum obtained after the risetime selection explained in text.}
\label{fig:RisetimeSpectrum}
\end{figure}

\medskip
Two other effects have been observed: one is a decrease in amplitude for events with long risetimes at low gains
(figure \ref{fig:RisetimeSpectrum}, left). This effect is caused by the ballistic effect, i.e.,
the time needed to collect all charge is comparable to the decharging time of the preamplifier.
To compensate this degradation, an upper limit in risetime of 13 $\mu$s has been applied. The second one is
a population of 22.1 keV events with lower amplitudes at short risetimes (same figure as before). This effect happens only
at the configuration without any field corrector. We have attributed these events to x-rays
depositing its energy near the ball, where a surface defect or a field distorsion may reduce the detector's gain.
To remove this contribution, a lower limit at risetime of 5 $\mu$s has been set.

\medskip
The resulting spectrum after the risetime selection is shown in figure \ref{fig:RisetimeSpectrum} (right).
The K$_{\alpha}$ (22.1 keV) and K$_\beta$ lines (24.9 keV) from Ag fluorescence are clearly distinguished,
as well as the fluorescence lines of the vessel (crome at 5.5 keV, iron at 6.4 keV) and
the source container (copper at 8.0 keV), and the corresponding Argon escape peaks (E.P.).
The peak at 22.1 keV is used to obtain the gas gain and energy resolution, fitting it to a gaussian funtion (red line).
The energy resolution of 6.4 keV line is around 23\% FWHM for all optimum cases and configurations, near the expected value
(20\% FWHM) derived from the 22.1 keV peak. The difference between both values could be explained by fitting errors
due to the presence of other peaks in the same range of energy.

\section{Results without the field corrector}
\begin{figure}[htb!]
\centering
\includegraphics[width=75mm]{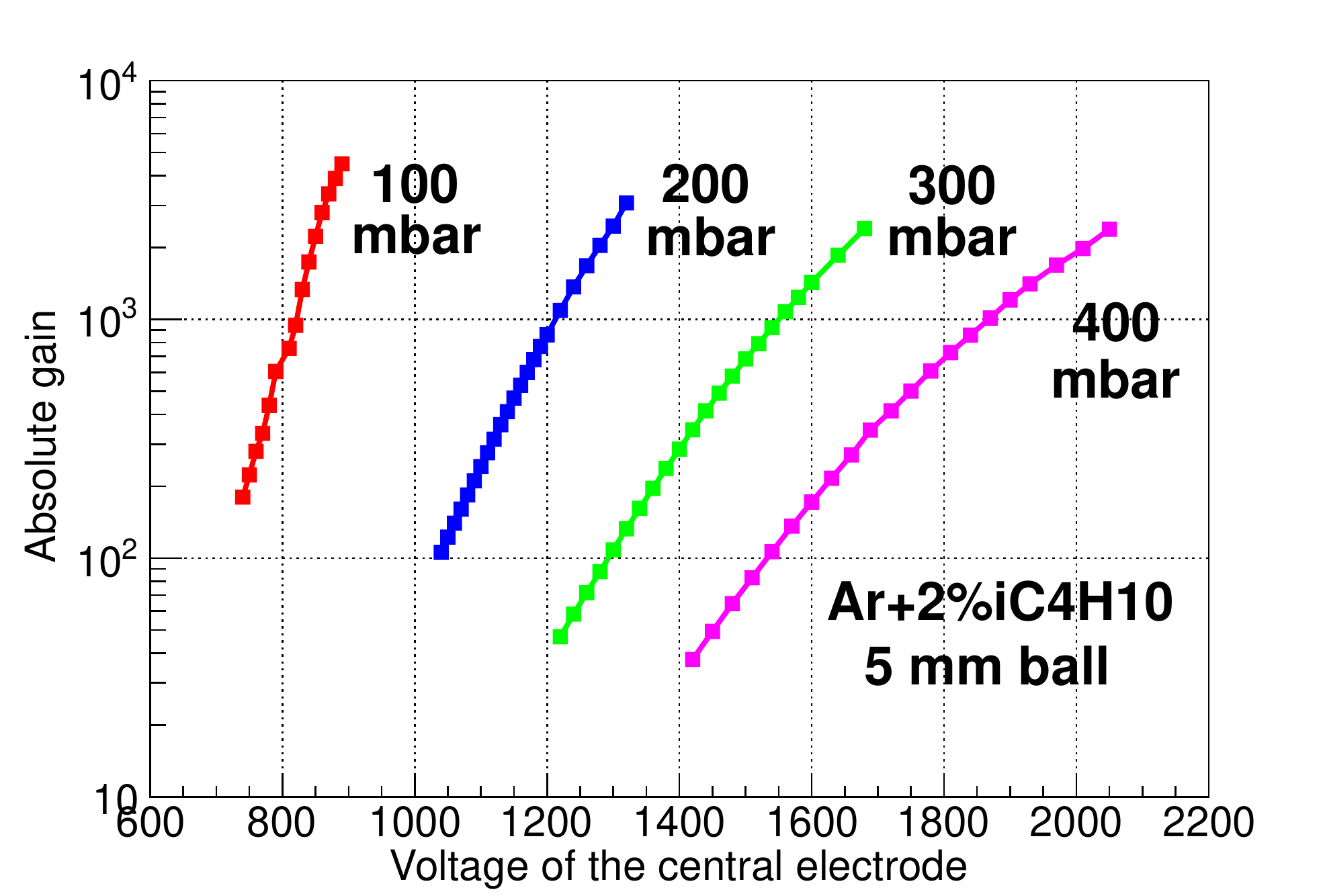}
\includegraphics[width=75mm]{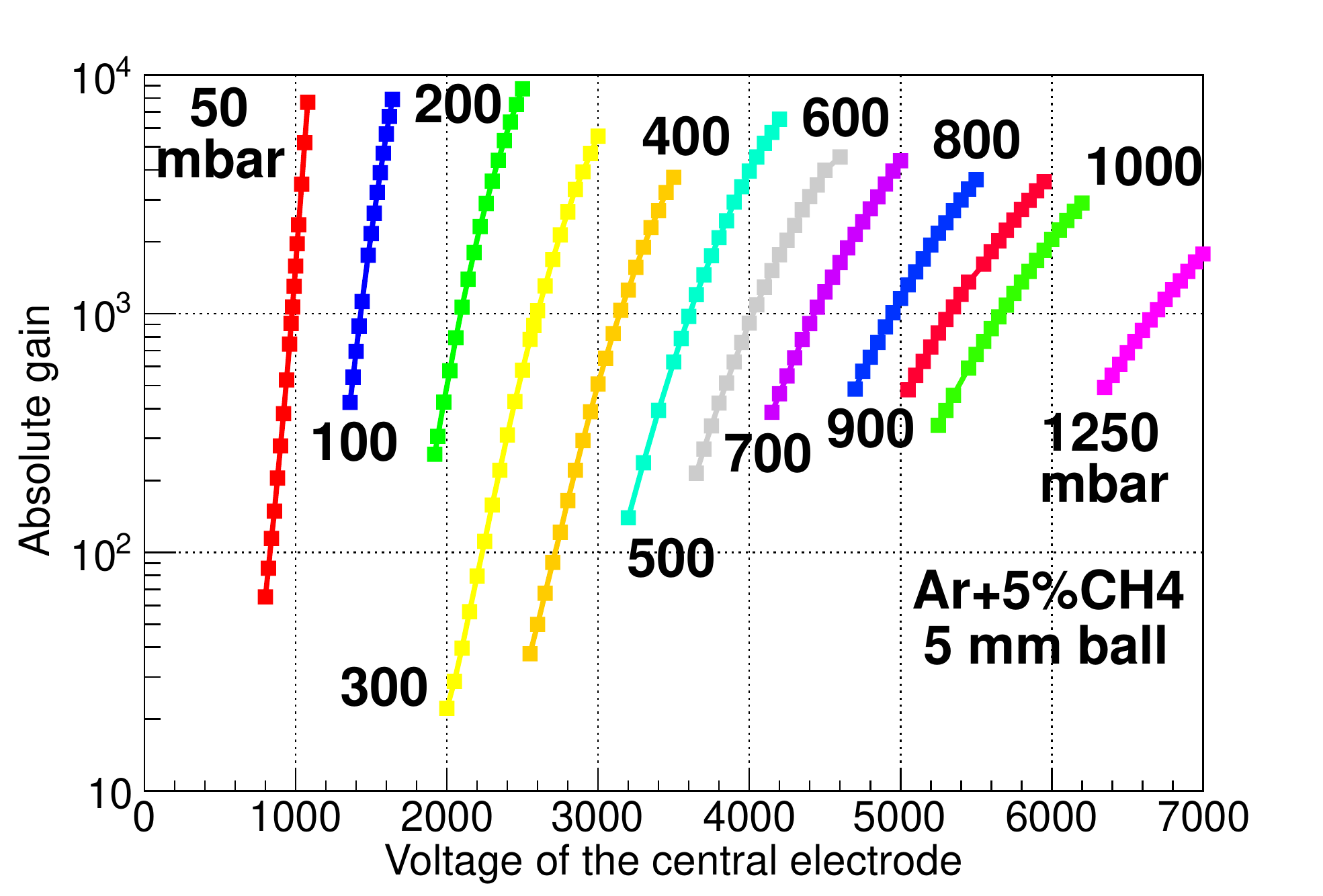}
\caption{Dependence of the absolute gain with the voltage of the central electrode
for Ar+2\% isobutane (left) and Ar+5\% methane (right) and gas pressures between 50 and 1250 mbar.}
\label{fig:GainArgon}
\end{figure}

The dependence of the peak position with the voltage generates the gain curves, shown in figure \ref{fig:GainArgon}
for both argon-based mixtures. The gain reaches values higher than $2 \times 10^3$ in all cases.
Moreover, the spark limit was not observed in any case and the data-taking was resumed
when the energy resolution degraded. The dependence of the gain with the voltage follows the expected exponential relation
based on the Rose-Korff model \cite{Rose:1941mr}, except for high voltages where gain tends to saturation.
We have attributed this effect to ion-backflow \cite{Colas:2004pc}, i.e, ions produced in the avalanche enter in the drift
volume and produce distorsions due to space-charge effects.

\begin{figure}[htb!]
\centering
\includegraphics[width=75mm]{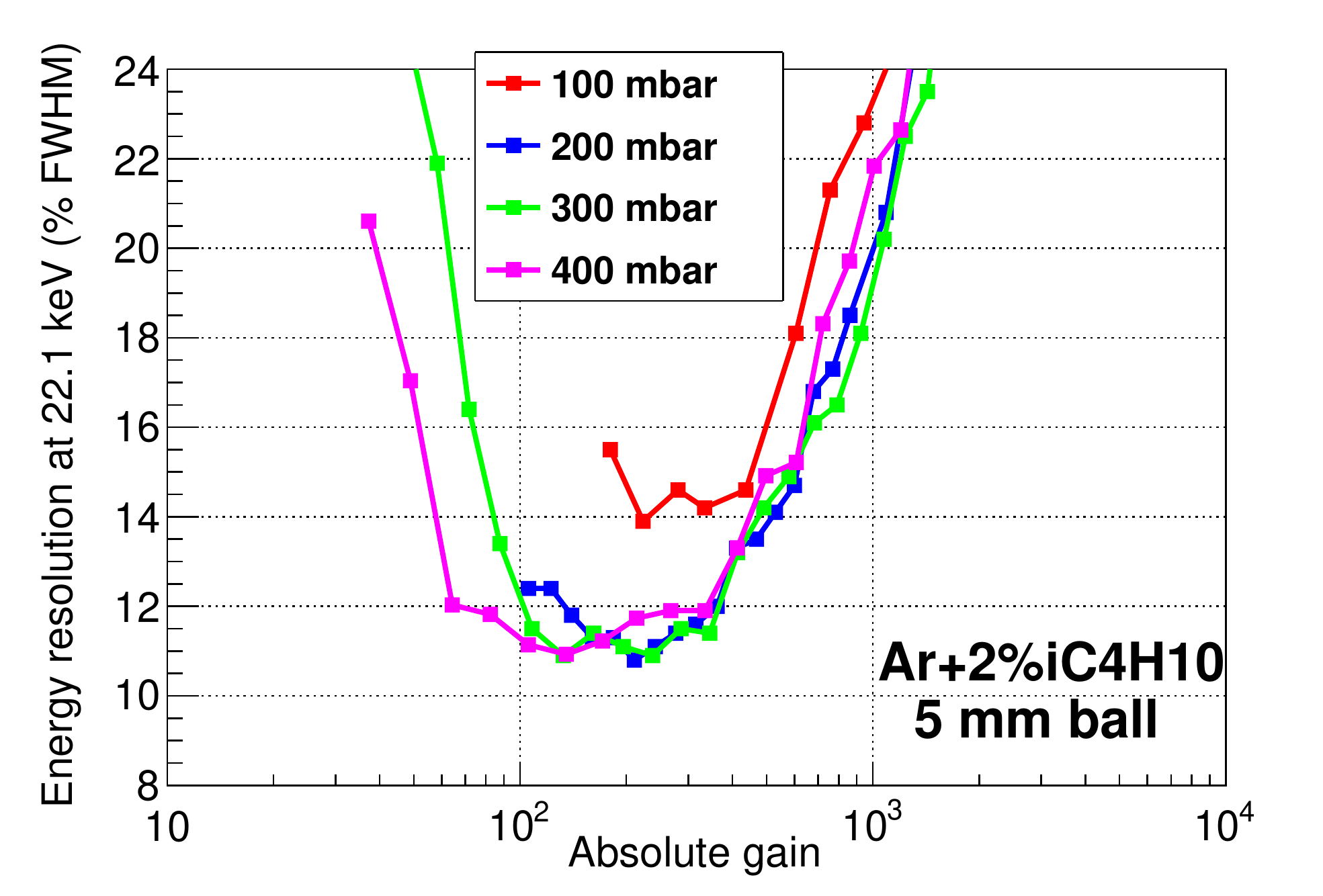}
\includegraphics[width=75mm]{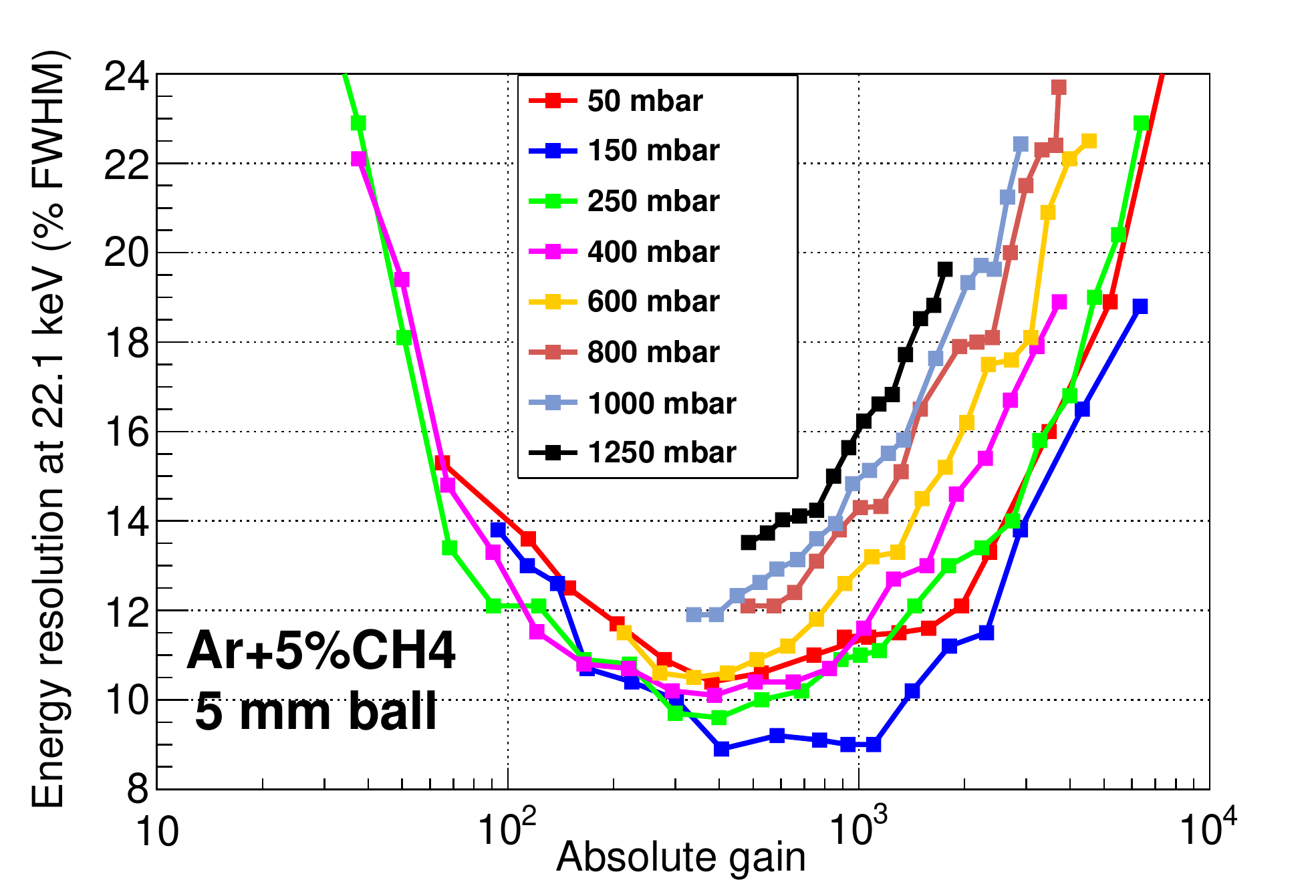}
\caption{Dependence of the energy resolution (\% FWHM) at 22.1 keV with the absolute gain
for Ar+2\% isobutane (left) and Ar+5\% methane (right) and gas pressures between 50 and 1250 mbar.}
\label{fig:EResArgon}
\end{figure}

\medskip
The energy resolution depends on the detector's gain, as shown in figure \ref{fig:EResArgon}.
At low values, the energy resolution degrades because the signal is comparable to noise.
As the gain increases, it reaches its best value and stays constant during a range of gains,
worsening again at higher gains. The surprising fact is how different are the gains where
this last degradation starts: $\approx 3 \times 10^2$ for Ar+2\%iC$_4$H$_{10}$
and $\approx 10^3$ for Ar+5\%iCH$_4$, i.e., a factor 3 better for methane
even if isobutane quenchs better the photons generated in the avalanche \cite{Agrawal:1988pca}.
For this reason, we can not attribute this effect to an increase of gain fluctuations
or the proximity to the spark limit, but to an effect probably correlated with the former gain saturation.
Just as a comparison, energy resolution starts degrading in argon-based mixtures
at $10^4$ for Micromegas detectors \cite{Iguaz:2012fi}.

\medskip
The best values obtained for energy resolution at 22.1 keV are 9\% FWHM in Ar+5\% methane at 150 mbar
and 11\% FWHM in Ar+2\% isobutane at 200-400 mbar, as shown in figure \ref{fig:EResArgon}. As noted before,
better values were expected for isobutane as this gas is a better quencher. 
These values are similar to those reported in \cite{Bougamont:2012eb}.
Finally, we have observed a degradation of energy resolution with pressure for the argon-methane mixture.
Its physical origin is still unknown but similar tendencies have been reported
for Micromegas detectors in CF$_4$ \cite{Jeanneret:2003pj} and Xe-TMA \cite{Cebrian:2012sc}.

\section{First results of the field corrector}
\begin{figure}[htb!]
\centering
\includegraphics[width=75mm]{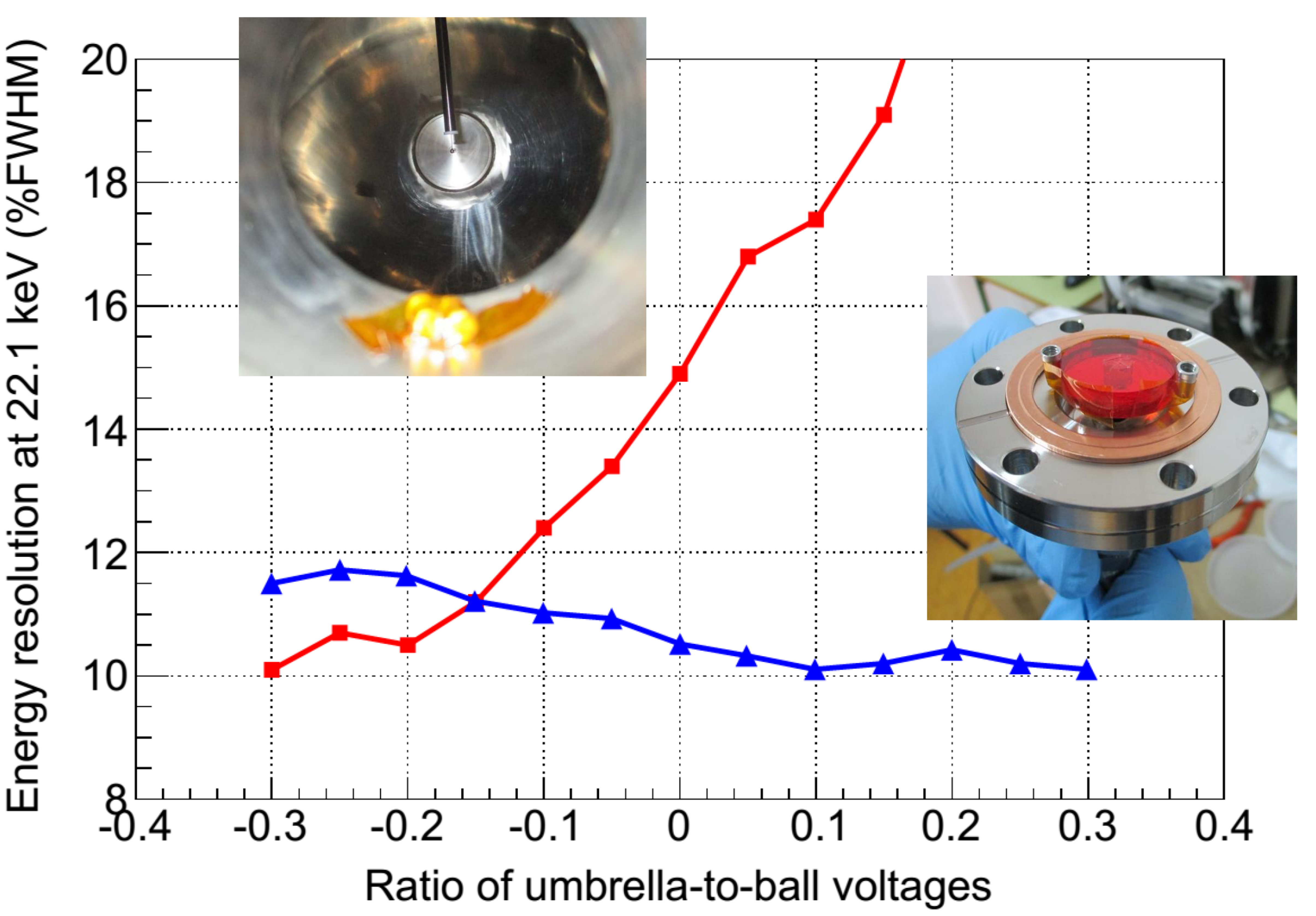}
\includegraphics[width=75mm]{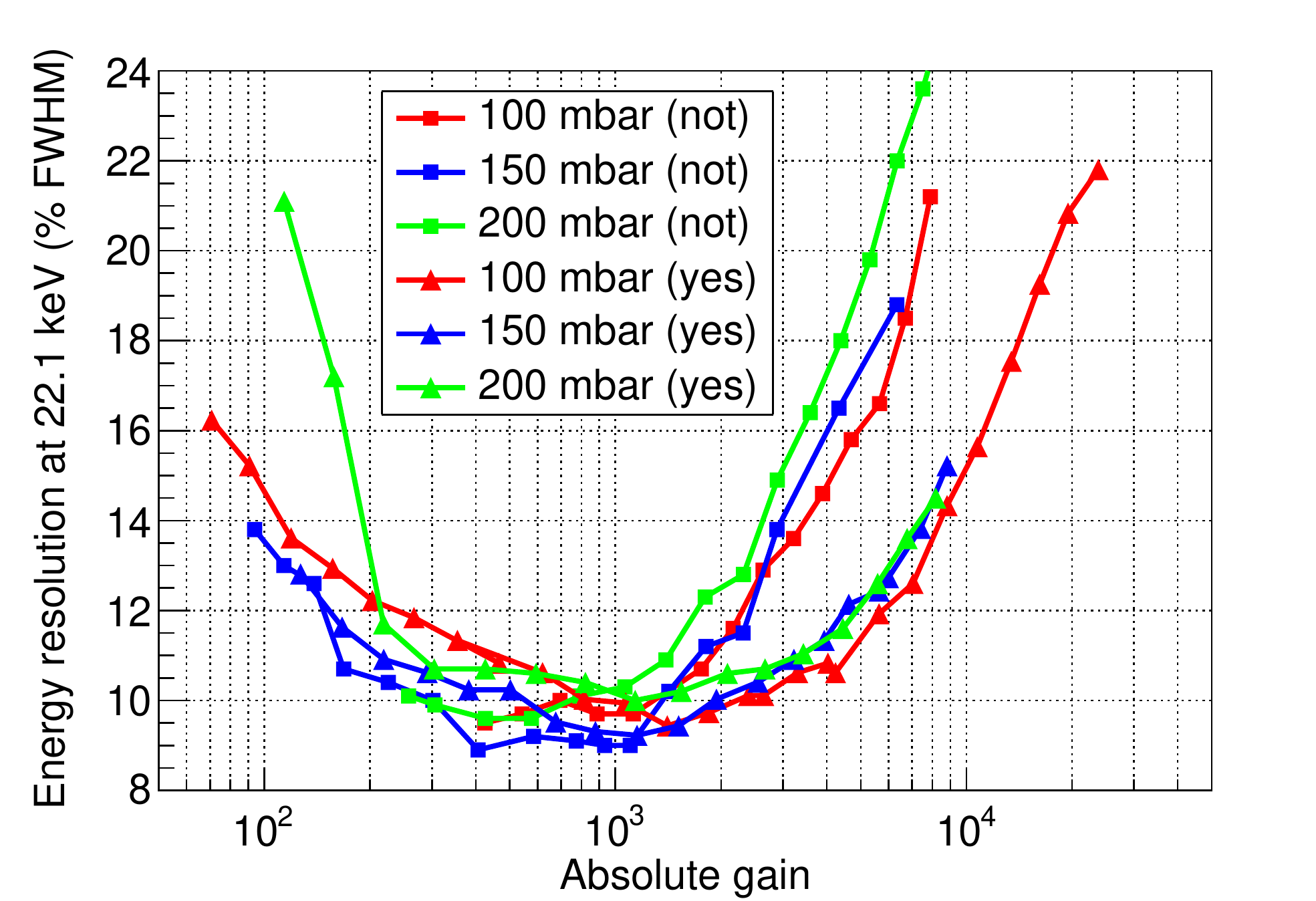}
\caption{Left: Dependence of the energy resolution (\% FWHM) at 22.1 keV with the ratio of umbrella-to-ball voltages
for the electrode configuration with the field corrector calibrating the SPC by a $^{109}$Cd source alternatively
situated inside the vessel at the lateral bore (red squares) and the bottom part (blue trianges). Data has been taken
in Ar+5\% methane at 100 mbar and setting the central ball at 1350 V. The two figures at the plot
show how the $^{109}$Cd source was installed.
Right: Dependence of the energy resolution (\% FWHM) at 22.1 keV with the absolute gain in Ar+5\% methane and
gas pressures between 100 and 200 mbar for the configuration without (squares) and with the field corrector (triangles).}
\label{fig:FieldCorrector}
\end{figure}

The effect of the umbrella field corrector (figure \ref{fig:Setup}, top-right) was studied calibrating the SPC
alternatively from the lateral bore and the bottom bore (blue stars at figure \ref{fig:Setup}, left)
in Ar+5\% methane at pressures between 100 and 200 mbar. The gain increased if negative voltages were applied
to the umbrella for both cases but no maximum was found. In terms of energy resolution
(figure \ref{fig:FieldCorrector}, left), we observed a clear improvement for lateral
calibrations if the field corrector was set to negative values, reaching an optimum at a ratio of -0.2. However,
the energy resolution slightly degraded for bottom calibrations. As a compromise, we fixed the ratio of
umbrella-to-ball voltages to -0.2 when generating gain curves. Then,
we replaced the lateral cap by an aluminized mylar window of 3.5 $\mu$m thickness,
and verified that the gain was uniform in all radial directions,
because the vessel's fluorescence lines of a long run (see figure \ref{fig:RTDisc}) were clearly defined.

\medskip
The absolute gain was a factor 5 higher in this configuration, i.e., a lower voltage must be applied to reach the
same gain. Apart from that, the degradation observed in energy resolution at high gains appeared at higher values
(figure \ref{fig:FieldCorrector}, right), which allowed us to reach an energy threshold below 1 keV
(figure \ref{fig:RTDisc}, right). This value is still far from the 0.1 keV reported in \cite{Bougamont:2012eb} because
the DAQ used in this work has no trigger optimization. We expect to implement this type of electronics in future updates.
The best value for energy resolution was 9.5\% FWHM at 22.1 keV for 100 and 150 mbar, similar to the configuration
without the field corrector.

\section{Conclusions and prospects}
The Spherical Proportional Counter (SPC) is a novel type of radiation detector, with a low energy threshold
and good energy resolution. This detector has a wide range of applications
like Dark Matter searches, neutrino or neutron detection. To increase the sensitivity of this detector,
the scability of its good properties is crucial. For this purpouse, we present the characterization of
a 1 meter diameter SPC in two argon-based mixtures and for pressures up to 1250 mbar.
We have studied two rod configurations: without and with an umbrella field corrector.

\medskip
For the first configuration, gains as high as $2 \times 10^3$ have been reached.
Gain curves show a saturation effect at high values, which seems to be correlated with a degradation in energy resolution
and has been attributed to ion-backflow.
The best values obtained for the energy resolution at 22.1 keV are 9\% FWHM in Ar+5\% methane at 150 mbar and
11\% FWHM in Ar+2\% isobutane at 200-400 mbar respectively for gains $(0.2-1.0) \times 10^3$ and $(0.1-0.2) \times 10^3$.
The energy resolution also degrades at high pressure but its physical origin is still unknown. In the second case,
the umbrella field corrector has made uniform the detector's gain in all radial directions.
The best value for energy resolution is 9.5\% FWHM at 100-150 mbar.
Energy resolution also degrades at high gains, but at higher values than in the first case, which has allowed us
to reduce the energy threshold below 1 keV. The new electrode is now being characterized for pressures up to 1250 mbar.

\medskip
In near term, we plan to update the acquisition system to further reduce the energy threshold and to study other light
gases like neon and helium. Our studies with these gases can be applied in the project SEDINE, developed by part of
the NEWS network. This project is operating a radiopure copper spherical vessel
at the Modane Underground Laboratory (LSM) since end 2012,
and aims at building a sphere of 2 meters of diameter filled with a light gas at 10-20 bar for Dark Matter searches.
This future experiment may reach sensitivities near DAMA signal for WIMPs masses as low as 1 GeV
if background level is around $10^{-2}$ keV$^{-1}$ kg$^{-1}$ day$^{-1}$ and the energy threshold
is kept below 0.1 keV for high pressures.
For further details, the reader is referred to \cite{Gerbier:2014gg, Dastgheibi:2014ad}.

\section*{Acknowledgements}
We are thankfull to our colleagues of the NEWS network for helpful discussions and encouragements.
We acknowledge support from the European Commission under the European Research Council
T-REX Starting Grant ref. ERC-2009-StG-240054 of the IDEAS program of the 7th EU Framework Program.
F.I. acknowledges the support from the \emph{Juan de la Cierva} program.

\end{document}